**Comment on Kuwahara *et al.*, Intensity Interference in a Coherent Spin-Polarized Electron Beam, Phys. Rev. Lett. 126, 125501 (2021)**

Herman Batelaan, Sam Keramati, T. J. Gay,

*Department of Physics and Astronomy, University of Nebraska-Lincoln, Lincoln, NE 68588, USA*

Recently, Kuwahara et al. [1] have reported the observation of a Hanbury Brown-Twiss electron antibunching dip (their Fig. 3) that is claimed to be the result of the Pauli exclusion principle. Such a feature can, however, be explained in other ways. Instead of resulting from (A), a Pauli blocking prohibiting more than one electron from populating a given free electron state, the dip could be caused by (B) the Coulomb force between two electrons deflecting them away from the coincidence detectors, or (C) a correlation between the electron spin states and source photoemission parameters.

In the earlier work of Kiesel et al. [2], causes A) and B) could not be distinguished and follow-up studies [3,4] found that the observed antibunching could be at least partially due to Coulomb repulsion. The exclusion of systematic effects that can mimic Pauli effects is difficult for the weak relative signals ($<10^{-3}$) associated with free electron anti-bunching experiments.

The antibunching dip profile reported in [1] does not carry any information in its shape other than that of the experimental timing resolution, in contrast to typical photonic or atomic antibunching experiments [5,6]. The reason is that their timing resolution (~100 ps) is much longer than their source's coherence time (~100 fs). The observed shape is thus governed by the detector response profile, and is characterizable by one number: the electron coincidence rate difference when switching between polarized and unpolarized electrons. This implies that the electron emission rate must be independent of any source emission parameters, lest an instrumental asymmetry (C) be mistaken for a Pauli effect (A).

In [1], the coincidence rate does decrease when the electrons are spin polarized. However, significant source effects can be expected when using strained GaAs photocathodes [7]. The electrons produced by photoemission from a standard (bulk) GaAs photocathode are polarized when the incident light producing them is either right- or left-hand circularly polarized and are unpolarized when linearly-polarized light is used. In the unstrained case, the heavy-hole and light-hole energy levels are degenerate at the Γ-point, leading to a maximum possible emitted electron polarization of 50%. Strain lifts the heavy-hole/light-hole degeneracy allowing 100% polarization to be produced, in principle, with a laser of sufficiently long wavelength to excite only the heavy-hole to conduction band transition. Linearly polarized light would yield no photoemitted electrons unless the laser has a short enough wavelength to bridge the light-hole/conduction-band gap, which is the case in the work of [1].

Unfortunately, ref. [1] does not discuss the respective emission rates for circular vs. linear light to the accuracy required (better than 1 part in $10^3$). This leaves open the question about whether the dip they observe is due to a variation in the quantum efficiency of their photocathode for linear vs. circular polarization. Moreover, since the strain on their GaAs/GaAsP sample is uniaxial, one would also expect a linear dichroism in the photoemission possibly as large as 15% [7] - much larger than the $10^{-3}$ reported effect. This reinforces the concern that an emission rate difference between circular light and linear light along either photocathode axis is present in these data. The authors do not say if they rotate the axis of the linearly-polarized light they use by 90° to check for such systematic effects.

Another pernicious source effect that can occur is a spatial variation of the laser beam focus on the photocathode correlated with polarization state due to the non-ideal nature of the liquid-crystal variable retarder (LCVR) [8]. Since the size of the laser focus used in this work was 2 microns, and negative electron affinity activation can lead to spatial variation of the quantum efficiency, the likelihood for systematic error in the measurement of a dip of the coincidence signal itself whose amplitude is $10^{-3}$ is potentially significant.

Thus it is important to consider how the electron beam intensity, and thus the experimental coincidence rate, depends on the photoemission laser's polarization. Ref. 1 provides only that the emission current is monitored to be 1.3 µA. This is not sufficient accuracy to settle this issue. For a study that claims to represent the first unambiguous observation of the HBT effect for free electrons, we hope that additional elements of the experiment will be reported that address the comments above.